\documentclass{ws-p9-75x6-50}

\begin{document}

\def\bm {{\mathop{\rm Mag}\nolimits }}
\def\bby  {{\bf y}}
\def\bs  {{\bf s}}
\def\bx  {{\bf x}}
\def\bxcl  {{\bf x}_{\rm cl}}
\def\bX  {{\bf X}}
\def\bXcl  {{\bf X}_{\rm cl}}
\def\det {\mathop{\rm det}\nolimits}
\def\jac {\mathop{\rm Jac}\nolimits}
\def\ru  {{\mbox{\rm u}}}
\def\rv  {{\mbox{\rm v}}}
\def\RR  {{\bf R}}
\newcommand{\bme}{ {\mbox{\boldmath $\eta$}} }

\title{Center of Light curves for Whitney Fold and Cusp}

\author{B. Scott Gaudi\footnote{Hubble fellow}}

\address{
School of Natural Sciences, 
Institute for Advanced Study, Princeton, NJ 08540,
USA\\E-mail: gaudi@sns.ias.edu}

\author{A. O. Petters}

\address{Department of Mathematics, Duke University, 
Science Drive, Durham, NC 27708, USA\\
E-mail: petters@math.duke.edu}  


\maketitle

\abstracts{
The generic, qualitative, local behavior of 
center-of-light curves near
folds and cusps are studied.
The results apply to any 
finite number of lens planes.}

\section{Center-of-Light Curves}

Let  $\bme: P \rightarrow S$
be a $k$-plane lensing
map, where $P$ is an open subset of the first lens plane
and $S$ the light source plane.
We define the {\it center-of-light} of a light source at $\bby \in S$ 
by:
\be
\label{def-colc}
\bxcl (\bby) = \frac{\sum_{\bx \in \bme^{-1} (\bby)} \  \bm (\bx; \bby)\  \bx}
                       {\sum_{\bx \in \bme^{-1} (\bby)} \  \bm (\bx; \bby)}, 
\ee
where 
$\bm (\bx; \bby) = 1/|\det[\jac \bme](\bx)|$ 
is the magnification of the lensed image
$\bx$ of $\bby$.  Suppose that the light source at $\bby$ moves
along a path $\bby (t)$.   Then the 
{\it shifted center-of-light curve} relative to the source's trajectory
is defined by
$\bXcl (t) = \bxcl (\bby (t)) - \bby(t).$
It is an important fact that locally stable $k$-plane lensing maps $\bme$
are generic and each is differentiably equivalent
about every critical point to either a Whitney fold or cusp
(Petters, Levine \& Wambsganss~\cite{plw}).
Whitney folds and cusps can then be used to study the
{\it generic, qualitative, local behavior} of $\bXcl$.

\begin{figure}[h!]
\centering
\epsfxsize=20pc 
\epsfysize=20pc 
\epsfbox{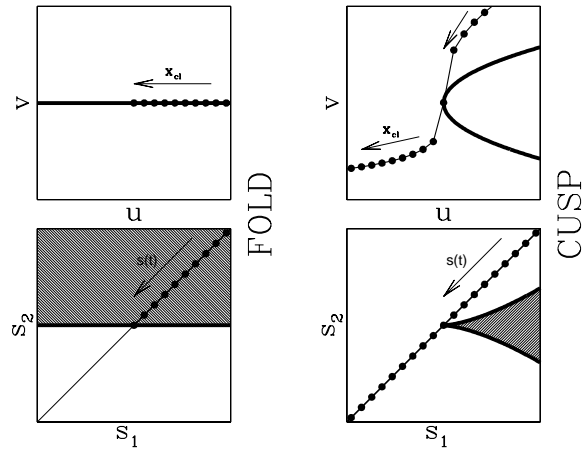}
\caption{Center-of-light curves for Whitney fold and cusp.}
\label{fig:center-of-light}
\end{figure}

A {\it Whitney fold} is a mapping 
$\bme_F: \RR^2 \rightarrow \RR^2$
of the form  $\bme_F (\ru, \rv) = (\ru, \rv^2).$
Let $(s_1, s_2)$ denote rectangular coordinates
on the target space of $\bme_F$.
The critical curve of $\bme_F$  is the $\ru$-axis,
while 
the caustic is the $s_1$-axis. 
For a light source at position $\bs = (s_1,s_2)$, we have
$\bme^{-1} (\bs) = \emptyset$ for $s_2 <0$
and 
$\bme^{-1} (\bs) = \{(s_1,\sqrt{s_2}), (s_1, -\sqrt{s_2})\}$
for $s_2 \ge 0$.   The lensed images have
magnification
$\bm ((s_1, \pm \sqrt{s_2}); \bs) = 1/(2\sqrt{s_2}).$ 
By (\ref{def-colc}), 
there is no center-of-light at
$\bs$ for $s_2 < 0$, while for
$s_2 \ge 0$ we have  
$\bxcl (\bs)  = (s_1, 0)$.  In other words, {\it the center-of-light
follows moves along the the critical curve}
(Fig.~\ref{fig:center-of-light}).
Suppose that the source follows a straight line
$\bs (t)  = t \exp(\theta_0)$, where 
$ -\infty < t < \infty$ and $\theta_0$ is
fixed with $0\le \theta_0 < 2 \pi$.    
Without loss of generality,  assume that the 
$\bs (t)$ is transverse to the caustic curve.
Then  
$\bXcl (t)$
moves along the $\rv$ axis, i.e.,
$\bXcl (t)  =   (0, s_2 (t))$ for $0 \le t < \infty$ 
and $\bXcl (t)$ has no values for $-\infty < t < 0.$

A {\it Whitney cusp} (or {\it pleat}) is a mapping
$\bme_F: \RR^2 \rightarrow \RR^2$
of the form  $\bme_F (\ru, \rv) = (\ru, -\ru \rv + \rv^3).$
The critical curve is a parabola,
$\ru = 3 \rv^2$, while the caustic
is the cusped curve $C (s_1, s_2) \equiv - (s_1/3)^3 
+ (s_2/2)^2  = 0$  with the origin a positive cusp.
Outside the caustic curve
(i.e., the region determined by $C(s_1, s_2) >0$),
a light source at $\bs = (s_1,s_2)$ has one lensed image,
namely, $(s_1, \rv (s_1,s_2))$,
where $\rv (s_1,s_2) = (s_2/2 \ + \ \sqrt{C(s_1,s_2)})^{1/3}
+ (s_2/2 \ - \ \sqrt{C(s_1,s_2)})^{1/3}.$
The magnification
is $\bm ((s_1, \rv (s_1,s_2));\bs) = 1/(-s_1 + 3 \rv^2(s_1,s_2)),$ 
which is positive. 
For a source trajectory
$\bs (t) = (s_1 (t),s_2 (t))$ lying outside the 
caustic curve, the  center-of-light moves along 
$\bxcl (t) = (s_1 (t), v (s_1 (t), s_2(t)))$
(Fig.~\ref{fig:center-of-light}).
The shifted center-of-light curve relative to
$\bs (t)$ 
is then given
by
$\bXcl (t)  = (0, \rv (s_1 (t) ,s_2 (t)) - s_2 (t)),$
which moves along 
the 
$\rv$-axis.  If 
the source moves along the negative $s_1$-axis
in the direction towards the cusp at the origin, 
say,
$\bs (t) = (s_1 (t), 0)$ with $s_1 (t) \le 0$,
then $\bXcl (t)  = (0,0)$ for all $t$ with
$s_1 (t) \le 0$.  In other words,  
{\it the shifted center of
light is located at the cusp critical point}  (i.e., the point
on
the critical curve that is mapped to the cusp
caustic point)
{\it and remains there
as the source moves
towards the cusp along} $(s_1 (t), 0)$.

\vskip-0.8in

The authors are preparing a thorough study of 
$\bxcl$ and $\bXcl$
near folds
and cusps.

\section*{Acknowledgments}

B.S.G. was supported in part by NASA through a Hubble Fellowship grant from the Space
Telescope Science Institute, which is operated by the Association of
Universities for Research in Astronomy, Inc., under NASA contract 
NAS5-26555.
A.P. 
was supported by an Alfred P. Sloan fellowship and
NSF CAREER grant DMS-98-96274.

\vfill

\end{document}